\documentclass[twocolumn,english,aps,prd,nofootinbib]{revtex4}
\usepackage[latin9]{inputenc}
\usepackage{xcolor}
\usepackage{babel}
\usepackage{amsmath}
\usepackage{amssymb}
\usepackage{cancel}
\usepackage{graphicx}
\usepackage{esint}
\usepackage[unicode=true]
 {hyperref}

\makeatletter

\newcommand{\lyxdot}{.}

\@ifundefined{textcolor}{}
{%
 \definecolor{BLACK}{gray}{0}
 \definecolor{WHITE}{gray}{1}
 \definecolor{RED}{rgb}{1,0,0}
 \definecolor{GREEN}{rgb}{0,1,0}
 \definecolor{BLUE}{rgb}{0,0,1}
 \definecolor{CYAN}{cmyk}{1,0,0,0}
 \definecolor{MAGENTA}{cmyk}{0,1,0,0}
 \definecolor{YELLOW}{cmyk}{0,0,1,0}
}

\usepackage{enumerate}

\makeatother

\begin{document}
\title{Beyond Schwarzschild--de Sitter spacetimes: I. A new exhaustive class
of metrics inspired by Buchdahl for pure $R^{2}$ gravity in a compact
form}
\author{Hoang Ky Nguyen}
\address{Baltimore, Maryland 21210, USA}
\email[\ \,Email:\ \ ]{HoangNguyen7@hotmail.com}

\date{September 28, 2022}
\begin{abstract}
\vskip2pt Some sixty years ago Buchdahl pioneered a program in search
of static spherically symmetric metrics for pure $\mathcal{R}^{2}$
gravity in vacuo (\emph{Nuovo Cimento, Vol 23, No 1, pp 141-157 (1962);
}\textcolor{purple}{\href{https://link.springer.com/article/10.1007/BF02733549}{https://link.springer.com/article/10.1007/BF02733549}}
\citep{Buchdahl-1962}). Surpassing several obstacles, his work culminated
in a non-linear second-order ODE which required being solved. However
Buchdahl deemed the ODE intractable and abandoned his pursuit for
an analytical solution. We have finally managed to overcome this remaining
hurdle and bring his program to fruition. \vskip6pt

Reformulating Buchdahl's ODE, we obtain a \emph{novel} class of metrics
(which we shall call the Buchdahl-inspired metrics hereafter) in a
compact and transparent expression:
\[
ds^{2}=e^{k\int\frac{dr}{r\,q(r)}}\left\{ p(r)\left[-\frac{q(r)}{r}dt^{2}+\frac{r}{q(r)}dr^{2}\right]+r^{2}d\Omega^{2}\right\} 
\]
in which the pair $\{p,q\}$ are two functions of the radial coordinate
$r$ obeying the evolution rules
\begin{align*}
{\displaystyle \frac{dp}{dr}}\, & =\,\frac{\ 3\,k^{2}}{4\,r}\frac{p}{q^{2}}\\
{\displaystyle {\displaystyle \frac{dq}{dr}}}\, & =\,{\displaystyle \Bigl(1-\Lambda\,r^{2}\Bigr)\,p}
\end{align*}
and the Ricci scalar is
\[
\mathcal{R}(r)=4\Lambda\,e^{-k\int\frac{dr}{r\,q(r)}}
\]

\noindent We are able to verify \emph{ex post}, via direct inspection,
that the metric given above satisfies the $\mathcal{R}^{2}$ vacuo
field equation
\[
\mathcal{R}\left(\mathcal{R}_{\mu\nu}-\frac{1}{4}g_{\mu\nu}\mathcal{R}\right)+\left(g_{\mu\nu}\square-\nabla_{\mu}\nabla_{\nu}\right)\mathcal{R}=0
\]
hence establishing its validity. The compact form above casts the
Buchdahl-inspired metric in a parallel resemblance with the classic
Schwarzschild--de Sitter (SdS) metric, with the case $k=0$ corresponding
to the SdS metric. \vskip6pt

We show why the Buchdahl-inspired metric, which exhibits non-constant
scalar curvature when $k\neq0$, defeats a ``no-go'' theorem proved
in \citep{Lust-2015-backholes} which posits that pure $\mathcal{R}^{2}$
gravity vacua are restricted to the Einstein spaces, $\mathcal{R}_{\mu\nu}=\Lambda g_{\mu\nu}$,
and the vanishing Ricci scalar spaces, $\mathcal{R}=0$. The aforementioned
``no-go'' theorem assumes a rapid asymptotic falloff for the metric
as $r\rightarrow\infty$. However, we find that the Buchdahl-inspired
metric evades that central assumption, which is overly restrictive.
\vskip6pt

A product of a fourth-derivative gravity, a Buchdahl-inspired metric
is specified by 4 parameters: $\,\Lambda$ measuring the scalar curvature
at largest distances, $k$ effecting the variation of the curvature
on the manifold, and $\{p_{0},q_{0}\}$ initiating the ``evolution''
of $\{p(r),q(r)\}$ along the radial direction, forming a two-dimensional
phase space. The class of Buchdahl-inspired metrics is \emph{exhaustive}
as it covers \emph{all} ``nontrivial'' static spherically symmetric
metrics admissible for pure $\mathcal{R}^{2}$ gravity in vacuo, with
the SdS metric being a special case, $k=0$. Transparently, the quartet
$\{\Lambda,\,k,\,p_{0},\,q_{0}\}$ spans a topological space with
all members in the class of Buchdahl-inspired metrics being \emph{smoothly
connected} to the SdS metrics when $k$ is continuously tuned to $0$.
In this respect, the Buchdahl-inspired metrics constitute a natural
enlargement suitably regarded as a framework ``beyond Schwarzschild--de
Sitter''. \vskip6pt

Our novel solution thereby completes Buchdahl's six-decades-old program.
We also explore the mathematical properties of the Buchdahl-inspired
metric in the limit of small $k$ and in the region around the coordinate
origin.
\end{abstract}
\maketitle

\section{\label{sec:Motivation}Motivation}

In a seminal paper entitled \emph{``On the Gravitational Field Equations
Arising from the Square of the Gaussian Curvature''} completed in
1961 \citep{Buchdahl-1962}, Hans A. Buchdahl pioneered -- yet left
unfinished -- a program to seek static spherically symmetric metrics
for \emph{pure} $\mathcal{R}^{2}$ gravity in vacuo, a theory that
excludes the Einstein-Hilbert term at the outset. Back in his time,
Buchdahl was motivated to consider the pure $\mathcal{R}^{2}$ action
as an interesting prototype for modified gravity. Recently, the quadratic
action has witnessed resurgence \citep{AlvarezGaume-2015,Lust-2015-backholes,Frolov-2009,Gurses-2012,Lu-2015,Nelson-2010,Pravda-2017,Stelle-2015};
one attractive feature of the pure $\mathcal{R}^{2}$ action is that
it is the only theory that is both ghost-free and scale invariant
\citep{Lust-2015-fluxes}.\vskip4pt

Despite making significant progress, unfortunately, Buchdahl discontinued
his efforts toward the finish line that was within striking distance.
The purpose of our current paper is to bridge the final remaining
gap in Buchdahl's ``abandoned'' program. The ultimate outcome is
a family of static spherically symmetric vacua, expressible in a compact
form, for the pure $\mathcal{R}^{2}$ action. We shall focus on the
mathematical aspects of these vacua in this paper, while leaving their
potential implications in physics for future research. \vskip4pt

As Buchdahl indicated therein \citep{Buchdahl-1962}, if one were
to adopt the canonical metric using Schwarzschild coordinates
\begin{align}
ds^{2} & =-A(r)\,dt^{2}+B(r)\,dr^{2}+r^{2}\,(d\theta^{2}+\sin^{2}\theta\,d\phi^{2})\label{eq:canonical}
\end{align}
then from the $\mathcal{R}^{2}$ vacuo field equation one would confront
a \emph{coupled} system of two equations for $A(r)$ and $B(r)$,
one of fourth- and one of third-differential orders. Eliminating one
of the two functions would yield a highly non-linear \emph{seventh-order}
ordinary differential equation (ODE).\vskip4pt

Nevertheless, Buchdahl was able to devise a judicious choice for the
metric alternative to \eqref{eq:canonical} such that the resulting
ODE -- albeit non-linear -- is only of \emph{second} differential
order \emph{which remained to be solved}. For the reader's convenience,
the original Buchdahl equation (as we shall call it as such, hereafter)
is
\begin{equation}
2t\,\frac{d^{2}q}{dt^{2}}+\left(\frac{1+t}{1-t}-\frac{3}{4}\,\frac{k^{2}}{q^{2}}\right)\frac{dq}{dt}=0\label{eq:Buchdahl-eqn}
\end{equation}
see Eqs. (1.7) and (3.4) in his original paper \citep{Buchdahl-1962}.
The metric he chose is then expressible in terms of the function $q(t)$
{[}N.B.: $t$ is not the time coordinate{]}, with the (Buchdahl) parameter
$k$ rendering Eq. \eqref{eq:Buchdahl-eqn} \emph{non-linear}.\vskip4pt

The Buchdahl equation \eqref{eq:Buchdahl-eqn} is very generic; it
captures \emph{all} ``nontrivial'' static spherically symmetric
vacua admissible for the pure $\mathcal{R}^{2}$ action, besides the
Einstein spaces (viz. $\mathcal{R}_{\mu\nu}=\Lambda\,g_{\mu\nu}$)
and the vanishing scalar curvature spaces (viz. $\mathcal{R}=0$).
Accordingly, \emph{if} an \emph{analytical} solution to the Buchdahl
equation can be found, then it would yield a powerful tool to tackle
the \emph{new physics} inherent in pure $\mathcal{R}^{2}$ gravity
\citep{Clifton-2011,deFelice-2010,Sotiriou-2008}. Crucially, as shall
be shown in this paper, the new (Buchdahl) parameter $k$\linebreak
in Eq. \eqref{eq:Buchdahl-eqn} would enable the $\mathcal{R}^{2}$
vacua to develop \emph{non-constant} scalar curvature.\vskip4pt

By and large, the Buchdahl equation was an impressive achievement.
Yet Buchdahl abandoned his pursuit for an analytical solution as he
judged his ODE intractable \footnote{\label{fn:Quote-Buchdahl}To quote Buchdahl from his original paper
(with notes in square brackets ours). In page 4 of \citep{Buchdahl-1962}:
\emph{``Unfortunately the simple appearance of {[}the non-linear
second order ODE{]} is deceptive. The best I have been able to achieve
is to obtain a solution in the form of a sequence of polynomials of
ascending powers of $t$.''} and in Page 8 of \citep{Buchdahl-1962}:
\emph{``{[}The ODE{]} does not appear to be soluble in terms of known
functions, nor does it appear to be reducible to a simpler form. It
therefore seems appropriate to determine a solution in ascending power
of $t$, or in some similar form.''}}.\linebreak This is an unfortunate twist of events as we find that
this is not the case \footnote{No further attempts either by Buchdahl or by others have been made
to solve his ODE since its publication.}. In this paper, we shall advance a number of mathematical maneuvers
to reformulate the Buchdahl equation \eqref{eq:Buchdahl-eqn} in a
more accessible form. From there, we are able to obtain a compact
expression for a new class of metrics which we shall call \emph{the
Buchdahl-inspired metrics}, thereby bringing his six-decades-old endeavor
to a successful outcome. 
\begin{center}
-----------------$\infty$-----------------
\par\end{center}

\vskip2pt For the reader's convenience, we shall briefly present
our result in what follows. The Buchdahl-inspired metric is neatly
expressible as\small
\begin{equation}
ds^{2}=e^{k\int\frac{dr}{r\,q(r)}}\left\{ p(r)\left[-\frac{q(r)}{r}dt^{2}+\frac{r}{q(r)}dr^{2}\right]+r^{2}d\Omega^{2}\right\} \label{eq:1.1}
\end{equation}
\normalsize with the Ricci scalar equal
\begin{equation}
\mathcal{R}(r)=4\Lambda\,\exp\left(-{\displaystyle k\int\frac{dr}{r\,q(r)}}\right)\label{eq:1.2}
\end{equation}
and the two auxiliary functions $p(r)$ and $q(r)$ evolving along
the radial direction $r$ per
\begin{align}
{\displaystyle \frac{dp}{dr}}\, & ={\displaystyle \,\frac{\ 3\,k^{2}}{4\,r}\frac{p}{q^{2}}}\label{eq:1.3}\\
{\displaystyle {\displaystyle \frac{dq}{dr}}}\, & =\,{\displaystyle \Bigl(1-\Lambda\,r^{2}\Bigr)\,p}\label{eq:1.4}
\end{align}
The deliberate resemblance of Eq. \eqref{eq:1.1} to a Schwarzschild--de
Sitter (SdS) metric makes the meaning of terms transparent. The compact
form \eqref{eq:1.1}--\eqref{eq:1.4} automatically encompasses the
constant-curvature SdS when $k$ equal zero \footnote{A fact to be shown in Sec. \ref{sec:Recovering-SdS}.},
in which case the non-linear and singular relation in \eqref{eq:1.3}
stays silent. A non-zero $k$, however, would trigger an interplay
between $p$ and $q$ via \eqref{eq:1.3} and \eqref{eq:1.4}, in
which case the Buchdahl-inspired metric acquires a \emph{non-constant}
scalar curvature per \eqref{eq:1.2}, potentially offering a host
of intricate phenomenology and \emph{new physics}.\vskip2pt
\noindent \begin{center}
-----------------$\infty$-----------------
\par\end{center}

\textcolor{black}{Our paper is organized as follows. In Sec. \ref{sec:Recasting-Buchdahl}
we shall rework Buchdahl's original paper in a simplified and straightforward
approach. Our two-fold aim is to derive the results directly from
the $\mathcal{R}^{2}$ field equation, and to arrive at an ODE which
is more generic than his original ODE. In Sec. \ref{sec:Our-short-cut}
we shall introduce a short-cut towards the (generalized) Buchdahl
equation while circumventing his original Hamiltonian-based procedure.
In Sec. \ref{sec:Our-solution} we shall cast his equation in a more
transparent way, then obtain a compact solution describing the new
class of Buchdahl-inspired metrics. In Sec. \ref{sec:Verification}
we shall outline the verification process which confirms the validity
of our Buchdahl-inspired metrics. Between Secs. \ref{sec:Recovering-SdS}
and \ref{sec:Degeneracy}, we shall investigate the Buchdahl-inspired
metrics in 4 situations: (i) recovering the SdS metric at $k=0$,
(ii) deriving a new metric for the small-$k$ limit; (iii) probing
the behavior of the metrics around the coordinate origin; and (iv)
uncovering a degeneracy in the overall solution. Section \ref{sec:Evading-a-proof}
points out an overly restrictive assumption in a proof proposed in
\citep{Lust-2015-backholes} against the existence of non-constant
curvature metrics (and the class of Buchdahl-inspired metrics). Sec.
\ref{sec:Summary} summarizes our work.\vskip4pt}

\section{\label{sec:Recasting-Buchdahl}Generalizing the Buchdahl equation:
$\ $A more direct route}

\textcolor{black}{In his original work \citep{Buchdahl-1962} Buchdahl
followed an arduous route. He designed a new Lagrangian, as a ``surrogate''
to the pure $\mathcal{R}^{2}$ gravity action, then applied the variational
principle on it. With the benefits of hindsight, we shall rework Buchdahl's
formulation in a more straightforward manner. We shall start directly
from the $\mathcal{R}^{2}$ vacuo field equation, conduct the standard
calculations, and reach the }\textcolor{black}{\emph{generalized}}\textcolor{black}{{}
Buchdahl equation. We shall try to retain as much as possible Buchdahl's
notation for the reader's convenience.\vskip4pt}

Following Buchdahl's notation, the metric in spherical coordinate
is written in the form

\noindent 
\begin{eqnarray}
ds^{2} & = & -e^{\nu(r)}dt^{2}+e^{\lambda(r)}dr^{2}+e^{\mu(r)}d\Omega^{2}\label{eq:Buchdahl-coordinate}\\
d\Omega^{2} & = & d\theta^{2}+\sin^{2}\theta d\phi^{2}\nonumber 
\end{eqnarray}
The vacuo field equation in the pure $\mathcal{R}^{2}$ action is
\begin{equation}
\mathcal{R}\left(\mathcal{R}_{\mu\nu}-\frac{1}{4}g_{\mu\nu}\mathcal{R}\right)+(g_{\mu\nu}\,\square-\nabla_{\mu}\nabla_{\nu})\mathcal{R}=0\label{eq:R2-field-eqn}
\end{equation}
and the ``trace'' equation in vacuo is
\begin{equation}
\square\,\mathcal{R}=0\label{eq:R2-trace-eqn}
\end{equation}
Since $\mathcal{R}$ is a function of $r$ only, we have \footnote{Recall that for a scalar field $\phi$: $\nabla_{\mu}\nabla_{\nu}\phi=\partial_{\mu}\partial_{\nu}\phi-\Gamma_{\mu\nu}^{\lambda}\partial_{\lambda}\phi$.}
\begin{equation}
\nabla_{\mu}\nabla_{\nu}\mathcal{R}=\partial_{\mu}\partial_{\nu}\mathcal{R}-\Gamma_{\mu\nu}^{r}\,\partial_{r}\mathcal{R}
\end{equation}
The $tt$-, $\theta\theta$-, and $rr$- components of the vacuo field
equation \eqref{eq:R2-field-eqn} read\small
\begin{align}
\mathcal{R}_{tt}-\frac{1}{4}g_{tt}\mathcal{R} & =-\Gamma_{tt}^{r}\frac{\mathcal{R}'}{\mathcal{R}}\label{eq:00-eqn}\\
\mathcal{R}_{\theta\theta}-\frac{1}{4}g_{\theta\theta}\mathcal{R} & =-\Gamma_{\theta\theta}^{r}\frac{\mathcal{R}'}{\mathcal{R}}\label{eq:22-eqn}\\
\mathcal{R}_{rr}-\frac{1}{4}g_{rr}\mathcal{R} & =-\Gamma_{rr}^{r}\frac{\mathcal{R}'}{\mathcal{R}}+\frac{\mathcal{R}''}{\mathcal{R}}\label{eq:11-eqn}
\end{align}
\normalsize The relevant Christoffel symbols and components of the
Ricci tensors are\small
\begin{align}
\Gamma_{tt}^{r}\,e^{\lambda-\nu} & =\frac{\nu'}{2}\label{eq:Chris-00}\\
\Gamma_{\theta\theta}^{r}\,e^{\lambda-\mu} & =-\frac{\mu'}{2}\label{eq:Chris-22}\\
\Gamma_{rr}^{r} & =\frac{\lambda'}{2}\label{eq:Chris-11}
\end{align}
and
\begin{align}
\mathcal{R}_{tt}e^{\lambda-\nu} & =\frac{\nu''}{2}+\frac{\nu'^{2}}{4}-\frac{\nu'\lambda'}{4}+\frac{\nu'\mu'}{2}\\
-\mathcal{R}_{\theta\theta}e^{\lambda-\mu} & =-e^{\lambda-\mu}+\frac{\mu''}{2}+\frac{\mu'^{2}}{2}+\frac{\nu'\mu'}{4}-\frac{\lambda'\mu'}{4}\\
-\mathcal{R}_{rr} & =\frac{\nu''}{2}+\frac{\nu'^{2}}{4}+\mu''+\frac{\mu'^{2}}{2}-\frac{\nu'\lambda'}{4}-\frac{\lambda'\mu'}{2}\label{eq:R-rr}
\end{align}
\normalsize Furthermore, the Jacobian is
\begin{equation}
\sqrt{-g}\triangleq\sqrt{-\det g}=e^{\frac{\nu}{2}+\frac{\lambda}{2}+\mu}\sin\theta
\end{equation}
giving
\begin{equation}
\sqrt{-g}\,g^{rr}=e^{\frac{\nu}{2}-\frac{\lambda}{2}+\mu}\sin\theta
\end{equation}

\noindent The three functions $\nu(r),\ \lambda(r),\ \mu(r)$ are
subject to an arbitrary coordinate transform. Buchdahl made a \emph{judicious
choice} that
\begin{equation}
\mu(r)\equiv\frac{1}{2}\left(\lambda(r)-\nu(r)\right)\label{eq:Buchdahl-choice}
\end{equation}
thus making
\begin{equation}
\sqrt{-g}\,g^{rr}=\sin\theta
\end{equation}
The ``trace'' equation \eqref{eq:R2-trace-eqn} \footnote{Recall that for a scalar field $\phi$\emph{: }\textcolor{black}{\emph{$\square\phi=\frac{1}{\sqrt{-g}}\partial_{\mu}(\sqrt{-g}g^{\mu\nu}\partial_{\nu}\phi)$.}}}
\begin{equation}
\left(\sqrt{-g}\,g^{rr}\,\mathcal{R}'\right)'=0
\end{equation}
is vastly simplified to
\begin{equation}
\mathcal{R}''=0
\end{equation}
hence
\begin{equation}
\mathcal{R}=\Lambda+k\,r\label{eq:R-vs-r}
\end{equation}
in which $\Lambda$ and $k$ are 2 constants. If $k=0$ the Ricci
scalar is a constant everywhere. For $k\neq0$ the Ricci scalar deviates
from constancy.\vskip4pt

With Buchdahl's choice \eqref{eq:Buchdahl-choice}, the relevant Ricci
components become:\small
\begin{align}
\mathcal{R}_{tt} & =\frac{\nu''}{2}e^{\nu-\lambda}\label{eq:Rtt}\\
\mathcal{R}_{\theta\theta} & =1+e^{-\frac{\nu}{2}-\frac{\lambda}{2}}\Bigl(\frac{\nu''}{4}-\frac{\lambda''}{4}\Bigr)\label{eq:Rthetatheta}\\
\mathcal{R}_{rr} & =-\frac{\lambda''}{2}+\frac{\lambda'^{2}}{8}-\frac{3\nu'^{2}}{8}+\frac{\nu'\lambda'}{4}\label{eq:Rrr}
\end{align}
\normalsize From \eqref{eq:Chris-00}, \eqref{eq:R-vs-r}, \eqref{eq:Rtt}
the $tt$-equation \eqref{eq:00-eqn} reads:\small
\begin{equation}
\frac{\nu''}{2}\,e^{\nu-\lambda}+\frac{1}{4}e^{\nu}\bigl(\Lambda+kr\bigr)=-\frac{\nu'}{2}e^{\nu-\lambda}\frac{k}{\Lambda+kr}
\end{equation}
leading to
\begin{equation}
\nu''+\frac{k}{\Lambda+kr}\,\nu'+\frac{1}{2}\bigl(\Lambda+kr\bigr)\,e^{\lambda}=0\label{eq:b.1}
\end{equation}
\normalsize From \eqref{eq:Chris-22}, \eqref{eq:R-vs-r}, \eqref{eq:Rthetatheta}
the $\theta\theta$-equation \eqref{eq:22-eqn} reads:\small
\begin{align}
1+e^{-\frac{\nu}{2}-\frac{\lambda}{2}}\Bigl(\frac{\nu''}{4}-\frac{\lambda''}{4}\Bigr)-\frac{1}{4}e^{\frac{\lambda}{2}-\frac{\nu}{2}}\bigl(\Lambda+kr\bigr)\ \ \ \ \ \nonumber \\
=\Bigl(\frac{\lambda'}{4}-\frac{\nu'}{4}\Bigr)e^{-\frac{\nu}{2}-\frac{\lambda}{2}}\frac{k}{\Lambda+kr}
\end{align}
leading to
\begin{equation}
\lambda''-\nu''+\frac{k}{\Lambda+kr}\bigl(\lambda'-\nu'\bigr)+\bigl(\Lambda+kr\bigr)e^{\lambda}=4e^{\frac{\nu}{2}+\frac{\lambda}{2}}
\end{equation}
\normalsize which, combined with \eqref{eq:b.1}, becomes:
\begin{equation}
\lambda''+\frac{k}{\Lambda+kr}\,\lambda'+\frac{3}{2}\bigl(\Lambda+kr\bigr)\,e^{\lambda}=4e^{\frac{\nu}{2}+\frac{\lambda}{2}}\label{eq:b.2}
\end{equation}
From \eqref{eq:Chris-11}, \eqref{eq:R-vs-r}, \eqref{eq:Rrr} the
$rr$-equation \eqref{eq:11-eqn} reads:\small
\begin{equation}
-\frac{\lambda''}{2}+\frac{\lambda'^{2}}{8}-\frac{3\nu'^{2}}{8}+\frac{\nu'\lambda'}{4}-\frac{1}{4}e^{\lambda}\bigl(\Lambda+kr\bigr)=-\frac{\lambda'}{2}\frac{k}{\Lambda+kr}
\end{equation}
leading to
\begin{equation}
\lambda''-\frac{k}{\Lambda+kr}\,\lambda'+\frac{\Lambda+kr}{2}\,e^{\lambda}-\frac{\lambda'^{2}}{4}+\frac{3\nu'^{2}}{4}-\frac{\nu'\lambda'}{2}=0\label{eq:b.3}
\end{equation}
\normalsize Now, eliminating $\lambda''$ from Eqs. \eqref{eq:b.2}
and \eqref{eq:b.3}, we get: \small
\begin{equation}
2e^{\frac{\nu}{2}+\frac{\lambda}{2}}-\frac{k}{\Lambda+kr}\,\lambda'-\frac{\Lambda+kr}{2}\,e^{\lambda}-\frac{\lambda'^{2}}{8}+\frac{3\nu'^{2}}{8}-\frac{\nu'\lambda'}{4}=0\label{eq:b.4}
\end{equation}
\normalsize\vskip4pt

Next, we make the following coordinate change which is slightly \emph{different}
from Buchdahl in his original paper:
\begin{equation}
\Lambda+kr=\Lambda\,e^{kz}\label{eq:my-coord-change}
\end{equation}
The first and second derivatives acting on $r$ become:\small
\begin{align}
\frac{d}{dr} & =\frac{dz}{dr}\frac{d}{dz}=\frac{e^{-kz}}{\Lambda}\frac{d}{dz}\label{eq:op-1}
\end{align}
\begin{align}
\frac{d^{2}}{dr^{2}} & =\frac{dz}{dr}\frac{d}{dz}\left(\frac{e^{-kz}}{\Lambda}\frac{d}{dz}\right)\\
 & =\frac{e^{-kz}}{\Lambda}\left(-\frac{ke^{-kz}}{\Lambda}\frac{d}{dz}+\frac{e^{-kz}}{\Lambda}\frac{d^{2}}{dz^{2}}\right)\\
 & =\frac{e^{-2kz}}{\Lambda^{2}}\left(\frac{d^{2}}{dz^{2}}-k\frac{d}{dz}\right)
\end{align}
\normalsize upon which Eqs. \eqref{eq:b.1}, \eqref{eq:b.2}, \eqref{eq:b.4},
respectively, become:\small
\begin{equation}
\frac{e^{-2kz}}{\Lambda^{2}}\left(\nu_{zz}-k\nu_{z}\right)+\frac{ke^{-2kz}}{\Lambda^{2}}\nu_{z}+\frac{\Lambda}{2}e^{kz+\lambda}=0\label{eq:b.5}
\end{equation}
\begin{equation}
\frac{e^{-2kz}}{\Lambda^{2}}\left(\lambda_{zz}-k\lambda_{z}\right)+\frac{ke^{-2kz}}{\Lambda^{2}}\lambda_{z}+\frac{3\Lambda}{2}e^{kz+\lambda}=4e^{\frac{\nu}{2}+\frac{\lambda}{2}}\label{eq:b.6}
\end{equation}
\begin{align}
\frac{ke^{-2kz}}{\Lambda^{2}}\lambda_{z}+\frac{\Lambda}{2}e^{kz+\lambda}+\frac{e^{-2kz}}{8\Lambda^{2}}\lambda_{z}^{2}\ \ \ \ \ \ \ \ \ \ \nonumber \\
-\frac{3e^{-2kz}}{8\Lambda^{2}}\nu_{z}^{2}+\frac{e^{-2kz}}{4\Lambda^{2}}\nu_{z}\lambda_{z} & =2e^{\frac{\nu}{2}+\frac{\lambda}{2}}\label{eq:b.7}
\end{align}
hence giving
\begin{equation}
\nu_{zz}+\frac{\Lambda^{3}}{2}e^{3kz+\lambda}=0\label{eq:b.8}
\end{equation}
\begin{equation}
\lambda_{zz}+\frac{3\Lambda^{3}}{2}e^{3kz+\lambda}=4\Lambda^{2}e^{2kz+\frac{\nu}{2}+\frac{\lambda}{2}}\label{eq:b.9}
\end{equation}
\begin{align}
\lambda_{z}^{2}-3\nu_{z}^{2}+2\nu_{z}\lambda_{z}+8k\lambda_{z}+4\Lambda^{3}e^{3kz+\lambda}\ \ \ \ \ \ \ \nonumber \\
=16\Lambda^{2}e^{2kz+\frac{\nu}{2}+\frac{\lambda}{2}}\label{eq:b.10}
\end{align}
\normalsize Further define \small
\begin{align}
\nu & =-u+v-kz+\ln4\label{eq:b.11}\\
\lambda & =3u+v-3kz+3\ln4\label{eq:b.12}\\
\mu & =\frac{\lambda}{2}-\frac{\nu}{2}=2u-kz+\ln4\label{eq:b.13}
\end{align}
\normalsize from which, together with \eqref{eq:b.8}--\eqref{eq:b.10},
we obtain\small
\begin{align}
u_{zz} & =16\Lambda^{2}e^{u}\left(1-\Lambda e^{2u}\right)e^{v}\label{eq:b.14}\\
v_{zz} & =16\Lambda^{2}e^{u}\left(1-3\Lambda e^{2u}\right)e^{v}\label{eq:b.15}\\
u_{z}v_{z} & =16\Lambda^{2}e^{u}\left(1-\Lambda e^{2u}\right)e^{v}+\frac{3k^{2}}{4}\label{eq:b.16}
\end{align}
\normalsize If $\Lambda=1$, these equations would be equivalent
to Eqs. (3.1), (3.3), and (3.4) in Buchdahl's original paper \citep{Buchdahl-1962}.\vskip4pt

Let us recap: $\ $So far, we have obtained the three equations \eqref{eq:b.14}--\eqref{eq:b.16}
for two unknown functions $u(z)$ and $v(z)$. However, the three
equations are \emph{not} independent. Upon taking derivative with
respect to $z,$ Eq. \eqref{eq:b.16} yields
\begin{align}
 & u_{zz}v_{z}+u_{z}v_{zz}=\nonumber \\
 & \ \ \ \ \ \ \ \ 16\left(e^{u}-3\Lambda e^{3u}\right)e^{v}u_{z}+16\left(e^{u}-\Lambda e^{3u}\right)e^{v}v_{z}\label{eq:b.17}
\end{align}
which is \emph{trivially} satisfied by Eqs. \eqref{eq:b.14} and \eqref{eq:b.15}.
Therefore, the system is \emph{not} over-determined. We shall discard
Eq. \eqref{eq:b.15} while keeping Eqs. \eqref{eq:b.14} and \eqref{eq:b.16}
from now on.

\section{\label{sec:Our-short-cut}Our shortcut leading to the \emph{generalized}
Buchdahl equation}

Note that Eq. \eqref{eq:b.14} is of second differential order and
Eq. \eqref{eq:b.16} is of first differential order. Eliminating one
of the functions $u$ or $v$ would \emph{in principle} produce a
\emph{third} differential order ODE.\vskip4pt

To proceed, Buchdahl next exploited some clever analogy of Eqs. \eqref{eq:b.14}--\eqref{eq:b.16}
with a Hamiltonian dynamics. However, with the benefit of hindsight,
we have found a \emph{shortcut} to be presented in what follows.\vskip4pt

Define $q$ as a function of $u$:
\begin{equation}
q:=u_{z}\label{eq:b.18}
\end{equation}
giving
\begin{equation}
u_{zz}=q_{z}=q_{u}u_{z}=q_{u}\,q\label{eq:b.19}
\end{equation}
Also, by viewing $v$ as a function of $u$, we have
\begin{equation}
v_{z}=v_{u}u_{z}=v_{u}\,q\label{eq:b.20}
\end{equation}
Combining \eqref{eq:b.14} and \eqref{eq:b.19}, we get
\begin{equation}
q\,q_{u}=16\Lambda^{2}e^{u}\left(1-\Lambda e^{2u}\right)e^{v}\label{eq:b.21}
\end{equation}
Combining \eqref{eq:b.16}, \eqref{eq:b.18} and \eqref{eq:b.20},
we get
\begin{equation}
q^{2}v_{u}=16\Lambda^{2}e^{u}\left(1-\Lambda e^{2u}\right)e^{v}+\frac{3k^{2}}{4}\label{eq:b.22}
\end{equation}
Now, make a substitution
\begin{equation}
u=\ln x\label{eq:b.23}
\end{equation}
which leads to
\begin{align}
q_{u} & =\frac{q_{x}}{u_{x}}=x\,q_{x}\label{eq:b.24}\\
v_{u} & =\frac{v_{x}}{u_{x}}=x\,v_{x}\label{eq:b.25}
\end{align}
From Eqs. \eqref{eq:b.21} and \eqref{eq:b.22} we thus get
\begin{align}
q\,q_{x} & =16\Lambda^{2}\left(1-\Lambda x^{2}\right)e^{v}\label{eq:b.26}\\
q^{2}v_{x} & =16\Lambda^{2}\left(1-\Lambda x^{2}\right)e^{v}+\frac{3k^{2}}{4x}=q\,q_{x}+\frac{3k^{2}}{4x}\label{eq:b.27}
\end{align}
Differentiating Eq. \eqref{eq:b.26} with respect to $x$
\begin{equation}
q_{x}^{2}+q\,q_{xx}=16\Lambda^{2}\left(1-\Lambda x^{2}\right)e^{v}v_{x}-32\Lambda^{3}x\,e^{v}\label{eq:b.28}
\end{equation}
and rewriting it as
\begin{equation}
q_{x}^{2}+q\,q_{xx}=q\,q_{x}v_{x}-\frac{2\Lambda x\,q\,q_{x}}{1-\Lambda x^{2}}\label{eq:b.29}
\end{equation}
Substituting Eq. \eqref{eq:b.27} into the RHS of Eq. \eqref{eq:b.29}
\begin{equation}
q_{x}^{2}+q\,q_{xx}=q_{x}^{2}+\frac{3k^{2}q_{x}}{4x\,q}-\frac{2\Lambda x\,q\,q_{x}}{1-\Lambda x^{2}}\label{eq:b.30}
\end{equation}
which leads to
\begin{equation}
q_{xx}+\frac{2\Lambda x}{1-\Lambda x^{2}}\,q_{x}=\frac{3k^{2}}{4x\,q^{2}}\,q_{x}\label{eq:b.31}
\end{equation}
At $\Lambda=1$, it duly recovers
\begin{equation}
xq_{xx}+\left(\frac{2x^{2}}{1-x^{2}}-\frac{3k^{2}}{4q^{2}}\right)\,q_{x}=0\label{eq:b.32}
\end{equation}
which is precisely Eqs. (4.8) and (3.4) in Buchdahl's 1962 \emph{Nuovo
Cimento} paper \citep{Buchdahl-1962}. \vskip4pt

Remarkably, the resulting ODE is of \emph{second} (instead of third)
differential order. Finally, upon substituting $x:=\sqrt{t}$, Eq.
\eqref{eq:b.31} becomes:
\begin{equation}
2t\,q_{tt}+\left(\frac{1+\Lambda t}{1-\Lambda t}-\frac{3k^{2}}{4q^{2}}\right)q_{t}=0\label{eq:b.35}
\end{equation}
which, at $\Lambda=1$, recovers Eqs. (4.10) and (3.4) in Buchdahl's
paper \citep{Buchdahl-1962}.\vskip4pt

We shall call Eq. \eqref{eq:b.32} the \emph{generalized} Buchdahl
equation hereafter. Our next task is to make further progress with
this equation. 

\section{\label{sec:Our-solution}A new class of Buchdahl-inspired metrics}

As we alluded to in the Motivation, Buchdahl deemed that his non-linear
ODE \eqref{eq:b.35} -- although ``deceptively simple'' -- was
insoluble and irreducible to simpler forms. He discontinued his pursuit
for an analytical solution and instead sought a power-expansion solution;
see Footnote \ref{fn:Quote-Buchdahl} in our current paper for his
reasoning.\vskip4pt

We find that this is not the case. The task of this section is to
reformulate the generalized Buchdahl equation in a more transparent
way, via which the final metric can be attained. We shall consider
$\Lambda\in\mathbb{R}$ in general. It turns out that the generalized
Buchdahl ODE \eqref{eq:b.31} can be cast in a more convenient form
as

\begin{equation}
\frac{d}{dx}\left(\frac{q_{x}}{1-\Lambda x^{2}}\right)=\frac{3k^{2}}{4x\,q^{2}}\left(\frac{q_{x}}{1-\Lambda x^{2}}\right)\label{eq:b.38}
\end{equation}

Next, let us define a new function $p(x)$ per
\begin{equation}
p(x):=\frac{q_{x}}{1-\Lambda x^{2}}\label{eq:b.39}
\end{equation}
which, upon combining with \eqref{eq:b.38}, produces a set of two
coupled non-linear first-order ODEs:
\begin{align}
p_{x} & \,=\,\frac{3k^{2}}{4\,x}\,\frac{p}{q^{2}}\label{eq:b.40}\\
q_{x} & \,=\,\left(1-\Lambda x^{2}\right)p\label{eq:b.41}
\end{align}
In terms of $x$, the functions $u$ and $v$ are, using Eqs. \eqref{eq:b.23}
and \eqref{eq:b.26}
\begin{align}
e^{u} & =x\label{eq:b.42}\\
e^{v} & =\frac{q\,q_{x}}{16\Lambda^{2}\left(1-\Lambda x^{2}\right)}=\frac{q\,p}{16\Lambda^{2}}\label{eq:b.43}
\end{align}
and the functions $\nu$, $\lambda$, and $\mu$ are, using Eqs. \eqref{eq:b.11}--\eqref{eq:b.13}
\begin{align}
e^{\nu} & =e^{-u+v-kz+\ln4}=\frac{4}{\Lambda^{2}e^{kz}}\frac{q\,p}{16\,x}\label{eq:b.44}\\
e^{\lambda} & =e^{3u+v-3kz+3\ln4}=\frac{64}{\Lambda^{2}e^{3kz}}\frac{x^{3}q\,p}{16}\label{eq:b.45}\\
e^{\mu} & =e^{2u-kz+\ln4}=\frac{4}{e^{kz}}x^{2}\label{eq:b.46}
\end{align}
From \eqref{eq:my-coord-change} we have
\begin{equation}
dr=\Lambda e^{kz}dz\label{eq:b.47}
\end{equation}
and since we also know from \eqref{eq:b.18} and \eqref{eq:b.23}
that
\begin{equation}
q=u_{z}=\frac{du}{dx}\frac{dx}{dz}=\frac{1}{x}\frac{dx}{dz}\label{eq:b.48}
\end{equation}
which leads to
\begin{equation}
dz=\frac{dx}{x\,q}\label{eq:b.49}
\end{equation}
we thus have
\begin{equation}
dr=\Lambda e^{kz}\,\frac{1}{x\,q}\,dx\label{eq:b.50}
\end{equation}
The metric initially expressed in \eqref{eq:Buchdahl-coordinate}
becomes:\small
\begin{align}
ds^{2} & =-e^{\nu}dt^{2}+e^{\lambda}dr^{2}+e^{\mu}d\Omega^{2}\nonumber \\
 & =-\frac{pq}{4\Lambda^{2}e^{kz}x}dt^{2}+\frac{4pqx^{3}}{\Lambda^{2}e^{3kz}}\left(\frac{\Lambda e^{kz}}{xq}dx\right)^{2}\ \ \ \ \ \nonumber \\
 & \ \ \ \ \ \ \ \ \ \ \ \ \ \ \ \ \ \ \ \ \ \ \ \ \ \ \ \ \ \ \ \ +\frac{4x^{2}}{e^{kz}}d\Omega^{2}\\
 & =\frac{4}{e^{kz}}\left\{ \frac{p}{4}\left[-\frac{q}{4\,x}\frac{dt^{2}}{\Lambda^{2}}+\frac{4\,x}{q}dx^{2}\right]+x^{2}d\Omega^{2}\right\} \label{eq:b.51}
\end{align}
\normalsize Finally, using the notation of $r$ in place of $x$,
and making the following replacements
\begin{equation}
\begin{cases}
p & \rightarrow4p\\
q & \rightarrow4q\\
k & \rightarrow-4k\\
kz & \rightarrow-kz+\ln4\\
t & \rightarrow\Lambda t
\end{cases}\label{eq:b.51b}
\end{equation}
we arrive at the family of Buchdahl-inspired metrics presented below.
\pagebreak
\noindent \begin{center}
\textbf{\emph{The Buchdahl-inspired metrics}}
\par\end{center}

\noindent \small\vskip2pt
\begin{equation}
ds^{2}=e^{k\int\frac{dr}{r\,q(r)}}\biggl\{ p(r)\biggl[-\frac{q(r)}{r}dt^{2}+\frac{r}{q(r)}dr^{2}\biggr]+r^{2}d\Omega^{2}\biggr\}\label{eq:b.52}
\end{equation}
\normalsize in which the evolution rules are
\begin{align}
\frac{dp}{dr} & =\frac{3k^{2}}{4r}\frac{p}{q^{2}}\label{eq:b.52b}\\
\frac{dq}{dr} & =\left(1-\Lambda\,r^{2}\right)p\label{eq:b.52c}
\end{align}
and, using \eqref{eq:R-vs-r}, \eqref{eq:my-coord-change}, \eqref{eq:b.49}
and \eqref{eq:b.51b}, the Ricci scalar equals to
\begin{equation}
\mathcal{R}(r)=4\Lambda\,e^{-k\int\frac{dr}{r\,q(r)}}\label{eq:b.53}
\end{equation}
There are two \emph{separate }sets of metrics depending on the sign
of $\Lambda$:
\begin{itemize}
\item Asymptotically de Sitter: $\Lambda>0$ and $r\in[0,\Lambda^{-\frac{1}{2}}${]}
\begin{align}
\mathcal{R}(r) & ={\displaystyle 4\Lambda\,\exp\,\Bigl[k\int_{r}^{\Lambda^{-\frac{1}{2}}}\dfrac{dr'}{r'\,q(r')}\,\Bigr]}\label{eq:b.54}
\end{align}
\item Asymptotically anti-de Sitter: $\Lambda\leqslant0$ and $r\in[0,\infty)$
\begin{align}
\mathcal{R}(r) & =4{\displaystyle \Lambda\,\exp\,\Bigl[k\int_{r}^{\infty}\dfrac{dr'}{r'\,q(r')}\,\Bigr]}\label{eq:b.57}
\end{align}
\end{itemize}
In either case, the upper bound for the integral in $\mathcal{R}(r)$
is chosen such that, at the largest distance allowable, the Ricci
scalar converges to $4\Lambda$.\vskip4pt

Compatible with a fourth-derivative action, each metric is specified
by four parameters: $\Lambda$ (the large-scale curvature), $k$ (the
deviation from constant curvature), $p(r_{0})$ and $q(r_{0})$ at
a reference distance $r_{0}$. \vskip4pt

We shall tentatively call the class of metrics represented in (\ref{eq:b.52}--\ref{eq:b.57})
the Buchdahl-inspired metrics and the coordinate system $(t,r,\theta,\phi)$
used therein the Buchdahl coordinates. The Buchdahl-inspired metrics
are complete and \emph{exhaustive}. \emph{All} ``nontrivial'' static
spherically symmetric vacuo metrics in pure $\mathcal{R}^{2}$ gravity
fall under the umbrella of the Buchdahl-inspired metrics.

\section{\label{sec:Verification}Verifying our solution via direct inspection}

It is desirable to confirm \emph{ex post} that our solution expressed
in \eqref{eq:b.52}--\eqref{eq:b.53} obeys the $\mathcal{R}^{2}$
vacuo field equation. We shall carry out this due diligence exercise
via direct inspection. The task is nontrivial because of the cross
dependence between $p(r)$ and $q(r)$. Below is our maneuver. \vskip4pt

First, we consider the line element
\begin{equation}
ds^{2}=-e^{\nu(r)}dt^{2}+e^{\lambda(r)}dr^{2}+e^{\mu(r)}d\Omega^{2}
\end{equation}
in which, by virtue of \eqref{eq:b.52}\small
\begin{align}
\nu(r) & :=\ln\left(f(r)\,\frac{p(r)\,q(r)}{r}\right)\\
\lambda(r) & :=\ln\left(f(r)\,\frac{p(r)\,r}{q(r)}\right)\\
\mu(r) & :=\ln\left(f(r)\,r^{2}\right)
\end{align}
\normalsize We further equate
\begin{equation}
f(r):=\exp\left(k\int\frac{dr}{r\,q(r)}\right)
\end{equation}
while leaving $p(r)$ and $q(r)$ \emph{unspecified} at the moment.
\vskip4pt

The relevant Christoffel symbols and Ricci tensor components are given
in \eqref{eq:Chris-00}--\eqref{eq:R-rr}. We use the symbolic manipulator
MAXIMA ONLINE interface to compute these six components and the Ricci
scalar $\mathcal{R}$. They are found to contain $p(r)$ and $q(r)$
and their higher-differential order terms up to the fourth order.
\vskip4pt

Next, we specify
\begin{align}
p'(x) & =\frac{3k^{2}}{4\,r}\frac{p(r)}{q^{2}(r)}\\
q'(x) & =(1-\Lambda\,r^{2})\,p(r)\label{eq:ver-1}
\end{align}
then use MAXIMA ONLINE to compute $p''(r)$, $q''(r)$, $p'''(r)$,
$q'''(r)$, $p''''(r)$, $q''''(r)$ and express each of them solely
in terms of $p(r)$ and $q(r)$. We then substitute these quantities
into the Christoffel symbols, the Ricci tensor components, and the
Ricci scalar obtained above. Despite their cumbersome appearances,
after all the dust settles, MAXIMA ONLINE determines that 
\begin{align}
\mathcal{R}\left(\mathcal{R}_{tt}-\frac{1}{4}g_{tt}\mathcal{R}\right)+\Gamma_{tt}^{r}\mathcal{R}' & \equiv0\\
\mathcal{R}\left(\mathcal{R}_{\theta\theta}-\frac{1}{4}g_{\theta\theta}\mathcal{R}\right)+\Gamma_{\theta\theta}^{r}\mathcal{R}' & \equiv0\\
\mathcal{R}\left(\mathcal{R}_{rr}-\frac{1}{4}g_{rr}\mathcal{R}\right)+\Gamma_{rr}^{r}\mathcal{R}'-\mathcal{R}'' & \equiv0
\end{align}
\emph{identically}. In addition, it produces
\begin{equation}
\mathcal{R}(r)\equiv\frac{4\Lambda}{f(r)}\ \ \ \forall r
\end{equation}
These outcomes solidly validate that our solution given in \eqref{eq:b.52}--\eqref{eq:b.53}\emph{
satisfies} the $\mathcal{R}^{2}$ vacuo field equation.\vskip6pt

Our MAXIMA codes used for this section are available in \citep{Nguyen-2022-Verify}.
We must note that another researcher independently and successfully
verified our solution using \emph{Mathematica}; his working notebook
is accessible in the public domain \citep{Shurtleff-2022}.

\section{\label{sec:Recovering-SdS}Recovering Schwarzschild--de Sitter metric
as special case at $k=0$}

Consider a metric with constant curvature, $\mathcal{R}\equiv4\Lambda\ \ \forall r$.
This requires $k=0$ and, from \eqref{eq:b.52b}
\begin{equation}
\frac{dp}{dr}=0
\end{equation}
or $p=p_{0}\equiv1$ without loss of generality. Then, from \eqref{eq:b.52c},
we subsequently have
\begin{align}
\frac{dq}{dr} & =1-\Lambda r^{2}\\
q & =r-\frac{\Lambda}{3}r^{3}-r_{\text{s}}\\
\frac{q}{r} & =1-\frac{\Lambda}{3}r^{2}-\frac{r_{s}}{r}
\end{align}
with $r_{\text{s}}$ being a constant of integration. The metric in
\eqref{eq:b.52} becomes
\begin{equation}
ds^{2}=-\Bigl(1-\frac{\Lambda}{3}r^{2}-\frac{r_{\text{s}}}{r}\Bigr)dt^{2}+\frac{dr^{2}}{1-\frac{\Lambda}{3}r^{2}-\frac{r_{s}}{r}}+r^{2}d\Omega^{2}\label{eq:SdS-metric}
\end{equation}
which is nothing but the classic SdS metric. This result also means
that the SdS metric is the \emph{only} vacuo metric with \emph{constant}
curvature available in pure $\mathcal{R}^{2}$ gravity.\vskip4pt

A Buchdahl-inspired metric can be made \emph{arbitrarily} close to
the SdS metric by tuning the parameter $k$ to zero. Hence, the quartet
$\{\Lambda,k,p_{0},q_{0}\}$ spans a topological space where all members
in the space are smoothly connected to the $k=0$ member (namely,
the set of SdS metrics).

\section{\label{sec:Small-k-limit}The small $k$ limit}

For $k=0$ we already have the solution considered in the preceding
section:
\begin{align}
p(r) & \equiv1\text{ (without loss of generality)}\\
q(r) & =r-\frac{\Lambda}{3}r^{3}-r_{\text{s}}
\end{align}
Let us consider up to $\mathcal{O}(k)$
\begin{align}
p(r) & =1+\mathcal{O}(k)\label{eq:5B.3}\\
q(r) & =\left(r-\frac{\Lambda}{3}r^{3}-r_{\text{s}}\right)+\mathcal{O}(k)\label{eq:5B.4}
\end{align}
Plugging them into \eqref{eq:b.52b} leads to
\begin{equation}
\frac{dp}{dr}=\mathcal{O}(k^{2})
\end{equation}
which then means
\begin{equation}
p(r)=1+\mathcal{O}(k^{2})\label{eq:5B.6}
\end{equation}
Note that this expression is valid up to $\mathcal{O}(k^{2})$ instead
of merely $\mathcal{O}(k)$ as in \eqref{eq:5B.3}. Plugging \eqref{eq:5B.6}
into \eqref{eq:b.52c} yields
\begin{equation}
\frac{dq}{dr}=\left(1-\Lambda r^{2}\right)+\mathcal{O}(k^{2})
\end{equation}
then
\begin{equation}
q=\left(r-\frac{\Lambda}{3}r^{3}-r_{\text{s}}\right)+\mathcal{O}(k^{2})
\end{equation}
Once again, this expression is valid up to $\mathcal{O}(k^{2})$ instead
of merely $\mathcal{O}(k)$ as in \eqref{eq:5B.4}. The conformal
factor in the metric is thus
\begin{equation}
e^{k\int\frac{dr}{r\,q(r)}}=e^{k\int\frac{dr}{r^{2}\left(1-\frac{r_{\text{s}}}{r}-\frac{\Lambda}{3}r^{2}\right)}+\mathcal{O}(k^{3})}
\end{equation}
The metric in \eqref{eq:b.52} becomes:\small
\begin{align}
ds^{2} & =e^{k\int\frac{dr}{r^{2}\left(1-\frac{r_{\text{s}}}{r}-\frac{\Lambda}{3}r^{2}\right)}}\Biggl\{-\Bigl(1-\frac{r_{\text{s}}}{r}-\frac{\Lambda}{3}r^{2}\Bigr)\,dt^{2}+\nonumber \\
 & \ \ \ \ \ \ +\frac{dr^{2}}{1-\frac{r_{\text{s}}}{r}-\frac{\Lambda}{3}r^{2}}+r^{2}\,d\Omega^{2}\Biggr\}+\mathcal{O}(k^{2})\label{eq:new-metric}
\end{align}
\normalsize and the Ricci scalar is \small
\begin{equation}
\mathcal{R}=4\Lambda\,\biggl[\,1-k\int\frac{dr}{r^{2}\bigl(1-\frac{r_{\text{s}}}{r}-\frac{\Lambda}{3}r^{2}\bigr)}\,\biggr]+\mathcal{O}(k^{2})\label{eq:new-Ricci}
\end{equation}
\normalsize This new metric is valid up to $\mathcal{O}(k^{2})$,
and would be useful for physical situations with small $k$, i.e.,
with a weak deviation from constant scalar curvature. The metric is
determined by \emph{three} parameters $\Lambda$, $r_{\text{s}}$,
and $k$, each representing a length scale.\vskip4pt

At $\mathcal{O}(k^{2})$, the new metric \eqref{eq:new-metric} only
differs from the SdS metric \eqref{eq:SdS-metric} by the conformal
factor $e^{k\int\frac{dr}{r^{2}(1-\frac{r_{\text{s}}}{r}-\frac{\Lambda}{3}r^{2})}}$.
Note that the pure $\mathcal{R}^{2}$ action is \emph{not} subject
to the conformal symmetry. As a result, the conformal factor is a
physical quantity; it explicitly participates in the Ricci scalar
rendering the latter \emph{non-constant} as is evident in \eqref{eq:new-Ricci}.

\section{\label{sec:Behavior}Behavior of Buchdahl-inspired metric around
the coordinate origin}

\begin{figure}[!t]
\centering{}\includegraphics[scale=0.5]{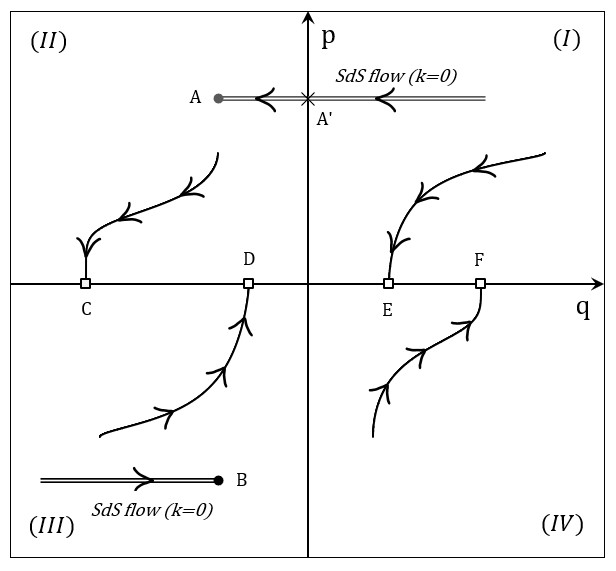}\caption{\label{fig:Buchdahl-flows}Evolution of $\{p(r),\,q(r)\}$ as $r$
approaches $0$. Points A and B are the end points of SdS flows (i.e.
$k=0$). Points C, D, E, and F are the end points of Buchdahl flows
($k\protect\neq0$), each starting from one of the four quadrants.}
\end{figure}

For any metric, the most interesting behavior should be around the
origin where singularities might occur. In the limit of $r\rightarrow0$,
the ``evolution'' rules \eqref{eq:b.52b} and \eqref{eq:b.52c}
become
\begin{align}
\frac{dp}{dr}\, & =\,\frac{3k^{2}}{4\,r}\frac{p}{q^{2}}\label{eq:5D.1}\\
\frac{dq}{dr}\, & \approx\,p\label{eq:5D.2}
\end{align}
The sign of $p$ solely determines the direction of flows for both
$p(r)$ and $q(r)$. Figure \ref{fig:Buchdahl-flows} shows the phase
space spanned by $\{p,\,q\}$ with $q$ the horizontal axis and $p$
the vertical axis. As $r$ moves \emph{toward }the coordinate origin,
Quadrants (I) and (II) correspond to monotonic decreasing $p$ and
$q$; Quadrants (III) and (IV) monotonic increasing $p$ and $q$.
Figure \ref{fig:Buchdahl-flows} shows the direction of the flow if
we start from a reference distance $r_{0}>0$ and move \emph{towards}
the origin. The SdS flows correspond to $k=0$ (thus, $p\equiv1$
and $q(r)=r-r_{\text{s}}+\frac{\Lambda}{3}r^{3}$ making ${\displaystyle \lim_{r\rightarrow0}q(r)=-r_{\text{s}}}$)
thus their end points belong to Quadrants (II) or (III). \vskip4pt

The horizontal axis is an attractor for all quadrants (note: we let
$r$ move \emph{toward} the coordinate origin). This can be shown
below.\vskip4pt

We shall let $p$ and $q$ converge to $p_{*}$ and $q_{*}$ when
$r\rightarrow0$ in the following manner
\begin{align}
p\, & \approx p_{*}+\bar{p}\,r^{\eta}\label{eq:5D.3}\\
q\, & \approx q_{*}+\bar{q}\,r^{\zeta}
\end{align}
 with $\eta>0$ and $\zeta>0$. First, let us assume $p_{*}\neq0$;
from \eqref{eq:5D.1}
\begin{equation}
\frac{dp}{dr}=\frac{3k^{2}}{4\,r}\,\frac{p}{q^{2}}\approx\frac{3k^{2}p_{*}}{4q_{*}^{2}}\,\frac{1}{r}
\end{equation}
making
\begin{equation}
p\,\approx\,-\frac{3k^{2}p_{*}}{4q_{*}^{2}}\,\frac{1}{r^{2}}+\text{const}
\end{equation}
which would diverge as $r\rightarrow0$ in contradiction with the
requirement \eqref{eq:5D.3}. Hence, $p_{*}$ must equal 0. This means
that \emph{every} trajectory must hit the horizontal axis as $r\rightarrow0$
from above. We shall only consider $q_{*}\neq0$ to this end. Since
$p_{*}=0$, the evolution rules \eqref{eq:5D.1} and \eqref{eq:5D.2}
become:
\begin{align}
\eta\,\bar{p}\,r^{\eta-1}\, & \approx\,\frac{3k^{2}}{4}\frac{\bar{p}}{q_{*}^{2}}\,r^{\eta-1}\\
\zeta\,\bar{q}\,r^{\zeta-1}\, & \approx\,\bar{p}\,r^{\eta}
\end{align}
giving
\begin{align}
\eta\, & =\,\frac{3k^{2}}{4q_{*}^{2}}>0\\
\zeta\, & =\,\eta+1>0\\
\bar{p}\, & =\,\zeta\,\bar{q}
\end{align}
Close to the origin, the functions thus are
\begin{align}
p(r)\, & \approx\,(\eta+1)\,\bar{q}\,r^{\eta}\\
q(r)\, & \approx\,q_{*}+\bar{q}\,r^{\eta+1}
\end{align}
The scalar curvature close to the origin behaves as
\begin{equation}
\mathcal{R}(r)\,\approx\,4\Lambda\,\exp\Bigl[\,-k\int\frac{dr}{r\,q_{*}}\,\Bigr]=\,4\Lambda\,r^{-\frac{k}{q_{*}}}
\end{equation}
As $r\rightarrow0^{+}$, the Ricci scalar vanishes or diverges depending
on the sign of $k/q_{*}$.\vskip6pt

As $r\rightarrow0^{+}$, the metric is approximately \small
\begin{equation}
ds^{2}\approx\,r^{\frac{k}{q_{*}}}\biggl\{(\eta+1)\frac{\bar{q}}{k}r^{\eta}\Bigl[-\frac{q_{*}}{kr}d\tilde{t}^{2}+\frac{kr}{q_{*}}dr^{2}\Bigr]+r^{2}d\Omega^{2}\biggr\}
\end{equation}
\normalsize which is specified by exactly \emph{three} parameters
$\{\Lambda,\,\frac{q_{*}}{k},\,\frac{\bar{q}}{k}\}$ with $\eta=\frac{3}{4}\left(\frac{k}{q_{*}}\right)^{2}$
and $\tilde{t}:=k\,t$.

\section{\label{sec:Degeneracy}A degeneracy in parameter space of Buchdahl-inspired
metric}

As the limit $k\rightarrow0$ corresponds to the SdS metric, we shall
consider only $k\neq0$ herein. If we make the following substitutions:
\begin{align}
q & :=k\,\tilde{q}\\
p & :=k\,\tilde{p}\\
t & :=k^{-1}\ \tilde{t}
\end{align}
 then the metric in \eqref{eq:b.52} becomes \small
\begin{equation}
ds^{2}=e^{\int\frac{dr}{r\,\tilde{q}(r)}}\left\{ \tilde{p}(r)\biggl[-\frac{\tilde{q}(r)}{r}d\tilde{t}^{2}+\frac{r}{\tilde{q}(r)}dr^{2}\biggr]+r^{2}d\Omega^{2}\right\} 
\end{equation}
\normalsize in which 
\begin{align}
\mathcal{R}(r)\, & =\,{\displaystyle 4\Lambda\,\exp\,\Bigl[\,-\int\dfrac{dr}{r\,\tilde{q}(r)}\,\Bigr]}\\
{\displaystyle \frac{d\tilde{p}}{dr}\ } & =\,\dfrac{3}{4\,r}\,\dfrac{\tilde{p}}{\tilde{q}^{2}}\\
{\displaystyle \frac{d\tilde{q}}{dr}\ } & =\,(1-\Lambda r^{2})\,\tilde{p}
\end{align}
Accordingly, despite being a product of a fourth-derivative action,
a Buchdahl-inspired metric is effectively characterized by only three
parameters. This degeneracy helps simplify the classification of Buchdahl-inspired
metrics. We shall carry out this task in a companion paper \citep{Nguyen-2022-Classify}.
\vskip4pt

Note that in Sec. \ref{sec:Small-k-limit} when treating the weak
non-constancy for the Ricci scalar, we made $k$ explicit. Nevertheless,
the metric obtained therein was specified by \emph{three} length scales
$\{\left|\Lambda\right|^{-\frac{1}{2}},\,r_{\text{s}},\,k\}$ in perfect
agreement with the number of degrees of freedom allowable by the degeneracy
uncovered in this section.\vskip4pt

\section{\label{sec:Evading-a-proof}How does Buchdahl-inspired metric circumvent
a \textquotedblleft proof\textquotedblright{} of nonexistence?}

In \citep{Lust-2015-backholes} Kehagias \emph{et al} sought black
hole solutions for the pure quadratic action. Curiously, they omitted
the Buchdahl equation and consequently overlooked the new class of
Buchdahl-inspired metrics uncovered in our current paper. They considered
only the two ``automatic'' vacuo configurations: (i) the zero-Ricci-scalar
spaces, $\mathcal{R}=0$, and (ii) the Einstein spaces, $\mathcal{R}_{\mu\nu}=\Lambda\,g_{\mu\nu}$.
Therein, they offered a neat proof that apparently rules out the existence
of non-constant curvature metrics (to which Buchdahl-inspired metrics
belong). However, the class of Buchdahl-inspired metrics \emph{defeat}
their proof by evading its central assumption. Below is how it happens.
\vskip4pt

Let us first recap the essence of the proof of Kehagias \emph{et al}.
Their proof is a type of ``no-go'', stating that all admissible
$\mathcal{R}^{2}$ vacua must have constant scalar curvature. The
authors in \citep{Lust-2015-backholes} started with the trace equation
of the pure $\mathcal{R}^{2}$ action in vacuo 
\begin{equation}
\square\,\mathcal{R}=0
\end{equation}
For the following metric
\begin{equation}
ds^{2}=-\mu(r)dt^{2}+\frac{dr^{2}}{\nu(r)}+r^{2}d\Omega^{2}
\end{equation}
the trace equation takes the form \footnote{Recall that for a scalar field $\phi$: $\square\,\phi=\frac{1}{\sqrt{-g}}\partial_{\mu}\left(\sqrt{-g}\,g^{\mu\nu}\partial_{\nu}\phi\right)$}
\begin{equation}
\left(r^{2}\sqrt{\mu\nu}\,\mathcal{R}'\right)'=0
\end{equation}
This leads to
\begin{equation}
\left(r^{2}\sqrt{\mu\nu}\,\mathcal{R}'\mathcal{R}\right)'=\cancel{\left(r^{2}\sqrt{\mu\nu}\,\mathcal{R}'\right)'}\,\mathcal{R}+r^{2}\sqrt{\mu\nu}\,(\mathcal{R}')^{2}
\end{equation}
from which one obtains the following identity:
\begin{align}
\int_{0}^{\infty}dr\,r^{2}\sqrt{\mu\nu}\,(\mathcal{R}')^{2} & =\int_{0}^{\infty}dr\left(r^{2}\sqrt{\mu\nu}\,\mathcal{R}'\mathcal{R}\right)'\label{eq:identity-1}
\end{align}
The right-hand side of \eqref{eq:identity-1} can be cast into a three-volume
integral which then turns into a two-dimensional surface integral
at infinity by virtue of the Gauss-Ostrogradsky divergence theorem:
\footnote{Recall that in spherical coordinates, for $\phi(r)$ and $\vec{A}=A(r)\,\hat{r}$:
$\vec{\nabla}\phi=\partial_{r}\phi(r)\,\hat{r}$ and $\vec{\nabla}.\vec{A}=\frac{1}{r^{2}}\partial_{r}\left(r^{2}A(r)\right)$.
The 3D divergence theorem for a generic vector field $\vec{A}$: $\int_{V}d^{3}V\,\vec{\nabla}.\vec{A}=\oint_{S}d\vec{S}.\vec{A}$}

\begin{align}
 & \int_{0}^{\infty}dr\left(r^{2}\sqrt{\mu\nu}\,\mathcal{R}'\mathcal{R}\right)'\nonumber \\
 & \ \ \ \ \ \ =\frac{1}{4\pi}\int d\Omega\int_{0}^{\infty}dr\,r^{2}\,\vec{\nabla}\bigl(\sqrt{\mu\nu}\,\mathcal{R}(\vec{\nabla}\mathcal{R})\bigr)\\
 & \ \ \ \ \ \ =\frac{1}{4\pi}\int d^{3}V\,\vec{\nabla}\bigl(\sqrt{\mu\nu}\,\mathcal{R}(\vec{\nabla}\mathcal{R})\bigr)\\
 & \ \ \ \ \ \ =\frac{1}{4\pi}\oint_{S}d\vec{S}\sqrt{\mu\nu}\,\mathcal{R}(\vec{\nabla}\mathcal{R})\\
 & \ \ \ \ \ \ =\lim_{r\rightarrow\infty}r^{2}\sqrt{\mu\nu}\,\mathcal{R}\mathcal{R}'\label{eq:limit}
\end{align}

Now, the authors of \citep{Lust-2015-backholes} posited that \emph{if
$\mathcal{R}'$ falls to zero rapidly enough as large distances} then
the limit in \eqref{eq:limit} vanishes, making
\begin{equation}
\int_{0}^{\infty}dr\,r^{2}\sqrt{\mu\nu}\,(\mathcal{R}')^{2}=0\label{eq:identity-2}
\end{equation}
Because of the non-negativity of the left-hand side of \eqref{eq:identity-2},
this would force $\mathcal{R}'=0$ \emph{everywhere}. QED. \vskip8pt

However, Buchdahl-inspired metrics invalidate this very assumption:
their Ricci scalar decays \emph{not} as rapidly to warrant \eqref{eq:identity-2}.
As a counterexample, in Sec. \ref{sec:Small-k-limit} we obtained
a metric with the Ricci scalar behaving at large distances as, per
Eq. \eqref{eq:new-Ricci}:
\begin{equation}
\mathcal{R}\approx4\Lambda-\frac{4k}{r^{3}}
\end{equation}
making 
\begin{equation}
\mathcal{R}'\approx\frac{12k}{r^{4}}
\end{equation}
thence
\begin{equation}
\lim_{r\rightarrow\infty}\left|r^{2}\sqrt{\mu\nu}\,\mathcal{R}\mathcal{R}'\right|=\lim_{r\rightarrow\infty}\left|\frac{48\Lambda k}{r^{2}}\sqrt{\mu\nu}\right|=16\Lambda^{2}\left|k\right|\neq0\label{eq:non-zero}
\end{equation}
given that $\mu\simeq\nu\simeq1-\frac{\Lambda}{3}r^{2}$ as large
distances. In general, the growth in $\mu$ and $\nu$ balances out
the decay in $\mathcal{R}'$; the proof in \citep{Lust-2015-backholes}
overlooked this compensation effect. \vskip4pt

The non-zero value in \eqref{eq:non-zero} renders the ``no-go''
proof in \citep{Lust-2015-backholes} inapplicable for the Buchdahl-inspired
metric \footnote{As an aside comment, the proof in \citep{Lust-2015-backholes} was
not water-tight. It should also have handled the intricacy introduced
into the 3D divergence theorem by way of the curved \emph{space} (which
in general is not 3D Euclidean).}.\vskip8pt

Before closing this section, we must make two additional comments:
\vskip6pt

First, the ``no-go'' proof provided in \citep{Lust-2015-backholes}
was previously offered by Nelson for the $\mathcal{R}+\mathcal{R}^{2}+\mathcal{C}_{\mu\nu\rho\sigma}\mathcal{C}^{\mu\nu\rho\sigma}$
action \citep{Nelson-2010}.\linebreak Nelson's proof similarly relied
on an overly restrictive assumption on the asymptotic falloff for
$\mathcal{R}'$ as $r\rightarrow\infty$. \vskip4pt

Second, in a 2015 paper \citep{Lu-2015}, L\"u \emph{et al.} reported
the existence of further black hole solutions (above the Schwarzschild
solution) for the \emph{Einstein-Weyl} gravity, $\mathcal{R}+\mathcal{C}_{\mu\nu\rho\sigma}\mathcal{C}^{\mu\nu\rho\sigma}$,
viz. with the $\mathcal{R}^{2}$ term being suppressed. These solutions
-- albeit \emph{not} in an analytical form -- would be in defiance
of Nelson's ``no-go'' proof \citep{Nelson-2010}. The authors therein
\citep{Lu-2015} identified a (sign) error in Nelson's proof rendering
it inapplicable for the Einstein-Weyl gravity. However, these authors
did not refute Nelson's proof for the pure $\mathcal{R}^{2}$ gravity;
they did not point out the problem with the asymptotic falloff assumed
in Nelson's ``no-go'' proof which would have precluded the existence
of Buchdahl-inspired metrics, as we have shown in this section.

\section{\label{sec:Summary}Summary}

In this paper, we show that pure $\mathcal{R}^{2}$ gravity admits
nontrivial vacuo configurations beyond the vanishing Ricci scalar
spaces $(\mathcal{R}=0)$ and the Einstein space $(\mathcal{R}_{\mu\nu}=\Lambda g_{\mu\nu})$.
\vskip4pt

The new solutions are inherent in a program which Hans Buchdahl originated
circa 1962. In a seminal -- yet obscure -- \emph{Nuovo Cimento}
paper \citep{Buchdahl-1962}, Buchdahl set forth to seek static spherically
symmetric solutions for the pure $\mathcal{R}^{2}$ action. His work
culminated in a non-linear second-order ODE that \emph{remained to
be solved}. If a solution to his ODE can be found, then a complete
set of vacua for pure $\mathcal{R}^{2}$ gravity would be readily
obtained. \vskip4pt

Despite its importance and potential, the Buchdahl equation has largely
escaped the attention of the gravitation research community since
its inception. Among the mere 40+ publications that cited Buchdahl's
original \emph{Nuovo Cimento} work, none have attempted to solve his
ODE \footnote{Based on NASA ADS and InspireHEP citation-trackers.}.
In this paper, we have finally obtained a \emph{novel} set of compact
solutions to the Buchdahl equation, thereby accomplishing his six-decades-old
goal seeking nontrivial vacuo metrics for pure $\mathcal{R}^{2}$
gravity.\vskip6pt

\paragraph{Our main result:}

We reformulated Buchdahl's original work via a more straightforward
route starting directly from the $\mathcal{R}^{2}$ vacuo field equation;
we thus departed from Buchdahl's arduous route that used the variational
principle on a ``surrogate'' Lagrangian. Along the way, we introduced
a few shortcuts. We are able to arrive at a \emph{generalized} Buchdahl
equation in the form of a non-linear second-order ODE: 
\begin{equation}
\frac{d^{2}q}{dr^{2}}+\frac{2\Lambda r}{1-\Lambda r^{2}}\,\frac{dq}{dr}=\frac{3k^{2}}{4r\,q^{2}}\,\frac{dq}{dr}\label{eq:Buchdahl-eqn-1}
\end{equation}
This ODE embodies the four parameters, $\{\Lambda$, $k$, $q(r_{0})$,
$\frac{dq}{dr}|_{r=r_{0}}\}$, of the \emph{fourth-order} $\mathcal{R}^{2}$
theory.\vskip4pt

Next, in place of the second-order ODE \eqref{eq:Buchdahl-eqn-1},
we are able to recast it in terms of two coupled non-linear first-order
ODEs:
\begin{align}
{\displaystyle \frac{dp}{dr}}\, & ={\displaystyle \,\frac{3k^{2}}{4\,r}\,\frac{p}{q^{2}}}\label{eq:6.1}\\
{\displaystyle {\displaystyle \frac{dq}{dr}}}\, & =\,{\displaystyle \Bigl(1-\Lambda\,r^{2}\Bigr)\,p}\label{eq:6.2}
\end{align}
From here, we are able to express the final solution in a neat resemblance
to the SdS metric to make the terms transparent and self-explanatory.
The Buchdahl-inspired metrics are in a compact representation:

\vskip-8pt\small
\begin{equation}
ds^{2}=e^{k\int\frac{dr}{r\,q(r)}}\left\{ p(r)\left[-\frac{q(r)}{r}dt^{2}+\frac{r}{q(r)}dr^{2}\right]+r^{2}d\Omega^{2}\right\} \label{eq:6.3}
\end{equation}
\normalsize with the Ricci scalar equal \small
\begin{equation}
\mathcal{R}(r)=4\Lambda\,\exp\left({\displaystyle -k\int\frac{dr}{r\,q(r)}}\right)\label{eq:6.4}
\end{equation}
\normalsize As is generally expected from a fourth-order theory,
a Buchdahl-inspired metric is specified by 4 parameters: $\Lambda$
as the large-distance scalar curvature, the (Buchdahl) parameter $k$
controlling the deviation of the Ricci scalar from constancy, $\{p_{0},\,q_{0}\}$
initiating the ``evolution'' flow. \vskip6pt

\paragraph{Validity of our solution:}

To allay any doubts, in Sec. \ref{sec:Verification}, we verified
by \emph{direct inspection} that the metric given in \eqref{eq:6.1}--\eqref{eq:6.4}
obeys the $\mathcal{R}^{2}$ vacuo field equation
\begin{equation}
\mathcal{R}\left(\mathcal{R}_{\mu\nu}-\frac{1}{4}g_{\mu\nu}\mathcal{R}\right)+\left(g_{\mu\nu}\square-\nabla_{\mu}\nabla_{\nu}\right)\mathcal{R}=0
\end{equation}
hence establishing the validity of our solution. The verification
process will be detailed in \citep{Nguyen-2022-Verify}. Note that
another researcher also successfully carried out his own verification
of our results, with his \emph{Mathematica} notebook accessible in
the public domain \citep{Shurtleff-2022}.\vskip6pt

\paragraph{Circumventing a \textquotedblleft no-go\textquotedblright{} theorem:}

In \citep{Lust-2015-backholes} it was proved that pure $\mathcal{R}^{2}$
vacua were restricted to the vanishing Ricci scalar spaces, $\mathcal{R}=0$,
and the Einstein spaces, $\mathcal{R}_{\mu\nu}=\Lambda g_{\mu\nu}$.
This ``no-go'' proof, if it were correct, would rule out the existence
of vacua with non-constant scalar curvature. Since Buchdahl-inspired
metrics project non-constant scalar curvature, as is evident per \eqref{eq:6.4}
for $k\neq0$, we must identify the cause of the conflict. In Sec.
\ref{sec:Evading-a-proof} we found that the ``no-go'' proof in
\citep{Lust-2015-backholes} imposed a rapid asymptotic falloff for
the metric at largest distances. Buchdahl-inspired metrics, however,
evade this overly restrictive assumption, thereby being able to circumvent
the proof. \vskip6pt

\paragraph{Recovering the SdS metric at $k=0$:}

The case of $k=0$ corresponds to the SdS metric in which $p(r)$
can be set identically equal to 1 and $q(r)$ contains a Schwarzschild
radius; see Sec. \ref{sec:Recovering-SdS}.\vskip6pt

\paragraph{Properties of the Buchdahl-inspired metrics:}

We examined the metrics in three situations: (i) the small $k$ limit;
(ii) the region around the coordinate origin; and (iii) a degeneracy
in the parameter space of the metrics. These results are shown in
Sec.s \ref{sec:Small-k-limit}, \ref{sec:Behavior}, and \ref{sec:Degeneracy}
respectively. A thorough systematic study of the metrics shall be
provided in \citep{Nguyen-2022-Classify}.\vskip6pt

\paragraph{A framework \textquotedblleft beyond Schwarzschild--de Sitter\textquotedblright :}

The family of Buchdahl-inspired metrics (\ref{eq:6.1}--\ref{eq:6.4})
is \emph{exhaustive}: it covers all nontrivial static spherically
symmetric vacuo configurations admissible in pure $\mathcal{R}^{2}$
gravity. Its parameters $\{\Lambda,\,k,\,p_{0},\,q_{0}\}$ form a
topological space that encloses the constant-curvature SdS metrics
($k=0$) and smoothly connects each non-constant curvature member
to an SdS metric when $k$ is tuned to 0.\vskip6pt

The Buchdahl-inspired metrics thus constitute a\emph{ bona fide} enlargement
of the SdS metric. It offers a nontrivial example in the context of
$3+1$ higher-order gravity that encompasses the SdS metric yet --
at the same time -- \emph{transcends} it. Hence the Buchdahl-inspired
metrics embody a framework ``beyond Schwarzschild--de Sitter''.
\vskip6pt

In closing, the compact representation \eqref{eq:6.1}--\eqref{eq:6.4}
of the Buchdahl-inspired metrics should equip future researchers with
a powerful tool to explore \emph{new physics} in pure $\mathcal{R}^{2}$
gravity with relative ease.
\begin{acknowledgments}
I thank the anonymous referee for his/her highly constructive feedback
toward the improved manuscript. I thank Dieter L\"ust for his encouraging
remark regarding the appeal of my Buchdahl-inspired solutions and
their ability to escape a ``no-go'' theorem laid out in Ref \citep{Lust-2015-backholes}.
I further thank Richard Shurtleff for his deep technical insights,
Sergei Odintsov for his helpful feedback, and Timothy Clifton for
his supportive comments.
\end{acknowledgments}

\end{document}